\documentclass[sigconf]{acmart}

\usepackage{algorithm}
\usepackage{color}

\usepackage{microtype} 
\usepackage{cleveref}  
\usepackage{hyperref}  
\usepackage{amssymb,amscd}

\usepackage{amsmath}   
\usepackage{booktabs}  
\usepackage{multirow}  
\usepackage{graphicx}  
\usepackage{subfig}    
\usepackage[justification=centering]{caption}   

\usepackage{algorithmic}

\makeatletter
\renewcommand\footnoterule{\hrule\@width0.4\linewidth}
\makeatother

\usepackage{booktabs} 

\AtBeginDocument{%
  \providecommand\BibTeX{{%
    \normalfont B\kern-0.5em{\scshape i\kern-0.25em b}\kern-0.8em\TeX}}}

\copyrightyear{2019}
\acmYear{2019}
\setcopyright{acmcopyright}
\acmConference[IWQoS '19]{IEEE/ACM International Symposium on Quality of Service}{June 24--25, 2019}{Phoenix, AZ, USA}
\acmBooktitle{IEEE/ACM International Symposium on Quality of Service (IWQoS '19), June 24--25, 2019, Phoenix, AZ, USA}
\acmPrice{15.00}
\acmDOI{10.1145/3326285.3329073}
\acmISBN{978-1-4503-6778-3/19/06}

\begin{document}

\title[ATMPA: Attacking ML-Visualization Methods via AEs]{ATMPA: Attacking Machine Learning-based Malware Visualization Detection Methods via Adversarial Examples}


%


\author{Xinbo~Liu, Jiliang~Zhang*, Yaping~Lin, He~Li}
\thanks{*Corresponding author.}

\affiliation{%
  \institution{College of Computer Science and Electronic Engineering, Hunan University, Changsha, China}
  \institution{Provincial Key Laboratory of Trusted System and Networks in Hunan University, China}
  \institution{Correspond to: zhangjiliang@hnu.edu.cn}
 }


\renewcommand{\shortauthors}{Xinbo and Jiliang, et al.}

\begin{abstract}
Since the threat of malicious software (malware) has become increasingly serious, automatic malware detection techniques have received increasing attention, where machine learning (ML)-based visualization detection methods become more and more popular.
In this paper, we demonstrate that the state-of-the-art ML-based visualization detection methods are vulnerable to Adversarial Example (AE) attacks.
We develop a novel Adversarial Texture Malware Perturbation Attack (ATMPA) method based on the gradient descent and \textit{L}-norm optimization method, where attackers can introduce some tiny perturbations on the transformed dataset such that ML-based malware detection methods will completely fail.
The experimental results on the MS BIG malware dataset show that a small interference can reduce the accuracy rate down to 0\% for several ML-based detection methods, and the rate of transferability is 74.1\% on average.
\end{abstract}

\ccsdesc[500]{Security and privacy~Malware and its mitigation}
\ccsdesc[500]{Security and privacy~Software and application security}

\keywords{Adversarial Examples,~~Malware Detection,~~Image Texture,~~Machine Learning,~Data visualization.}



\maketitle


\section{Introduction}    
\label{Introduction}
With the rapid development of the Internet and smartphones, the number of malicious software has been growing unexpectedly in recent years.
More than 3.24 million new malwares were detected in Android application market in 2016 \cite{tam2017evolution}.
In 2017, Google Play intercepted more than 700,000 malicious software applications with an annual growth of 70\% \cite{google2017}.
These counterfeit applications use unwrapped content or Malware to impersonate legitimate applications and interfere with the normal software market.
According to the report from AV-TEST \cite{sr2017}, even though the popular ransomware accounts for less than 1\% of the total malware, billions of dollars have been lost since 2017.
Even worse, with the upcoming malicious programs (\textit{i.e.,} malware) and potential unwanted applications (\textit{e.g.,} PUA) \cite{sr2019}, the security situation becomes more serious than before.

Malware denotes a particular type of programs that perform malicious tasks to illegal control computer systems by breaking software processes to obtain the unauthorized access, interrupt normal operations and steal information on computers or mobile devices.
There is a variety of malwares including viruses, Trojans, worms, backdoors, rootkits, spyware and panic software \cite{makandar2017malware}.
Countermeasures against the malware threat can be classified into three cases: digital signatures \cite{nataraj2011malware}, static code analysis \cite{kang2015detecting} and dynamic code analysis \cite{huang2018adversarial}, to safeguard the digital world away from malware.
For traditional digital signature methods, since the number of new signatures released every year grows exponentially \cite{nataraj2011malware}, it is undesirable to use a digital signature-based method to determine each sample and detect the malicious behavior one by one.
Static code analysis can detect malware in a complete coverage through the disassembling, but it is not robust to the code obfuscation \cite{kancherla2013image}.
In addition, the executable files must be unpacked and decrypted before analysis, which significantly limits the detection efficiency.
Compared to the static code analysis, the dynamic code analysis does not need to unpack or decrypt execution files in a virtual environment.
However, it is time-intensive and resource-consuming, and scalability is another issue \cite{huang2018adversarial}.
In addition, methods mentioned above are difficult to detect some certain malicious behaviors that are well camouflaged or not satisfied by trigger conditions.

In recent years, researchers borrow the visualization idea from the field of computer forensics and network monitoring, and use machine learning (ML) to classify malware \cite{kancherla2013image,han2015malware,nataraj2011malware}.
By converting binary code into image data, theses ML-based visualization malware detection methods can not only visualize the features of various types of samples but also improve the detection speed for malicious software.
The detection process is more simplified than the conventional approaches.
Optimizations for ML algorithms have further improved the malware visualization detection techniques \cite{makandar2017malware,huang2017r2,grosse2017adversarial}, in terms of preventing zero-day attacks, non-destructive detection of malware without extracting preselected features.
Meanwhile, it improves the accuracy and robustness for detection and reducing time and memory consumption.

In this paper, we propose the first attack approach on ML-based visualization malware detection methods based on Adversarial Examples (AEs) \cite{aeZhang2019}, named Adversarial Texture Malware Perturbation Attack (ATMPA).
We extract the features of the state-of-the-art approaches for AE generation and use the Rectifier in neural-network hidden layers to improve the robustness for AEs training.
The ATMPA method is deployed into three malware visualization detectors to verify the attack effectiveness and transferability.
In the experimental evaluation, we use the malware dataset from the MS BIG database and convert it to a set of binary texture grayscale images.
The AEs generated by training such grayscale images are used to attack three ML-based detectors, i.e. CNN, SVM, RF and their wavelet combined versions.
The experimental results demonstrate that all of ML-based malware detection methods are vulnerable to our ATMPA attack and the success rate is 100\%.
The transferability test on different ML-based detection methods show that the ATMPA method can achieve an attack transferability up to 88.7\% (74.1\% on average).

The contributions of this paper are as follows:
\begin{itemize}
\item Propose the first framework that utilizes AEs to attack ML-based malware visualization detectors, and demonstrate the security vulnerabilities of ML applications.
\item Analysis of various AE variants to summarize an optimal infringement for ML-based malware detectors.
\item Evaluation and comparison analysis for the proposed ATMPA in terms of the effectiveness and transferability for CNN-, SVM- and RF-based malware detectors.
\item Extracting the feature of AE-based attacks along with several countermeasure solutions.
\end{itemize}

The rest of this paper is organized as follows: Section~\ref{Related work} details the related work of malware visualization detection methods associated with the AE techniques.
Three kinds of ML-based malware visualization detection algorithms are introduced in Section~\ref{Preliminary}.
Section~\ref{Methodology} describes the proposed ATMPA method.
We show the experimental results in Section~\ref{Experiment} and demonstrate the effectiveness of ATMPA.
Moreover, we give some qualitative defensive strategies against AE-based attacks in Section~\ref{Discussion}.
Finally, we make a conclusion and provide some future research directions in section~\ref{Conclusion}.

\section{Related work}
\label{Related work}
AEs are designed by attackers to fool deep learning models. Different from real examples, AEs can hardly be distinguished by human eyes, but mislead the model to predict incorrect outputs and therefore threaten security critical deep-learning applications \cite{aeZhang2019}.
 In general, transferability and robustness are two key features for AEs.
 The transferability refers to the degree to which a particular AE attack can be transferred to function as other types of AEs for different attacks.
 The robustness represents the ability to resist or overcome AE attacks.
 Prior work can be classified into un-targeted and targeted AEs.
 Un-targeted AEs can cause the classifier to produce any incorrect output without specifying a predicted category, while targeted AEs cause the classifier to produce a specific incorrect output.
 In the past three years, various AE generation methods were proposed, such as L-BFGS \cite{szegedy2013intriguing}, FGSM \cite{goodfellow2014explaining}, C\&W attack \cite{carlini2017towards} and DeepFool \cite{moosavi2016deepfool}.
 Among them, researches have shown that FGSM and C\&W attack methods are the popular choices for generating AEs \cite{aeZhang2019}.
 Hence we focus on the both in this paper.

 Code analysis-based malware detection can be classified into static code analysis and dynamic code analysis.
 Static code analysis is used to analyze malware without executing the program.
 The detection patterns used in static analysis include byte-sequence $n$-grams, string signature, control flow graph and opcode (operational code) frequency distribution \cite{gandotra2014malware}.
 On the other hand, analyzing the behaviour of a malicious code being executed in a controlled environment, such as virtual machine, simulator and emulator, is called as dynamic code analysis \cite{huang2018adversarial}.
 Dynamic code analysis is more effective as it discloses the natural behaviour of malware.
 Nowadays, there exists several automatic tools for dynamic analysis, such as TTAnalyzer \cite{bayer2006ttanalyze} and Norman Sandbox \cite{sandbox2010norman}.
 However, it is time-consuming to understand malware behavior by reading the analysis report generated by these tools.	

 In order to accelerate the malware detection process, ML-based visualization detection methods have received a great attention in recent years.
 This kind of method has been proposed for detecting and classifying unknown samples from either the known malware families or underlining those samples with unknown behaviour.
 However, ML-based methods have limited detection ability for different sample types.
 To overcome this issue, the ML-based visualization detection methods have been proposed.
 The indistinguishable malware can be easily detected by the detectors based on RF \cite{vsrndic2013detection}, SVM \cite{makandar2017malware}, CNN \cite{huang2017r2} by virtue of the transformed binary grayscale images.

 \begin{figure}[b]  
  \centering
  \includegraphics[scale=0.52]{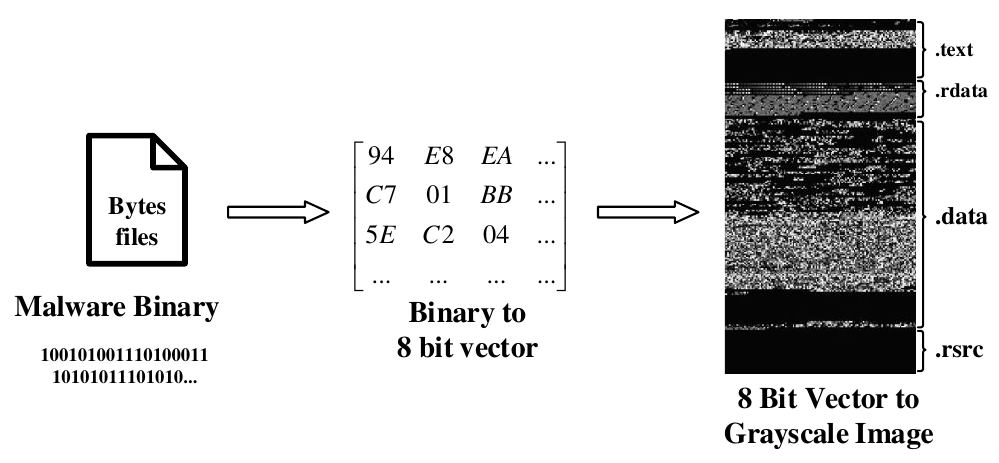}
  \caption{Malware visualization transformation}
  \label{fig:fig1}
 \end{figure}

 Unfortunately, AEs are used to bypass the detection of malicious codes.
 Gross \textit{et. al.} used AEs to interfere with the binary features of Android malware \cite{grosse2017adversarial}.
 They aimed to attack DNN-based detection for Android malware and retain malicious features in Apps.
However, such attack is only adapted to some Android malware samples processed by binary features.
 As for the end-to-end static code detection techniques using CNN, Kreuk \textit{et. al.} used AEs to extract the feature information from binary code and disguised malware as a benign sample by injecting a small sequence of bytes (payload) in the binary file \cite{kreuk2018fooling}.
 This method not only retains the original function of malware but also deceives the global binary malware detector.
 However, it is extremely sensitive to the discrete input dataset such as executable bytes.
 Minor modifications to the bytes of a file may lead to significant changes in its functionality and validity.
 Hu \textit{et. al.} proposed a method of generating sequential AEs by improving the robustness of its adversarial pertaining process, which can be used to attack the sequential API features-based malware in RNN detection system \cite{hu2017black}.
 The drawbacks of this method are time-consuming and large overhead in the whole processing.

\section{Preliminary}
\label{Preliminary}
 \subsection{Malware Visualization}
 \label{Pre MV}
 Malware visualization methods \cite{han2015malware,makandar2015overview,makandar2017malware} transform the hard-to-identify code information into image data with certain feature information.  
 A specific segment of code represents a definite type of malware.
 Lee \textit{et al.} firstly applied the visualization technology to accelerate the process of malware detection  \cite{lee2011study}.
 In the visualization process, we first collect a set of transformed malware samples and then read them as a vector of 8-bit unsigned integers.
 After that, we separate the set of vectors into a two-dimensional array.
 The corresponding value in the array can be expressed as the gray value of the image in the range from 0 to 255, where 0 indicates black and 255 means white.
 According to the different characteristics and analytical requirements for different data sets, the width of these transformed images could be appropriately adjusted as needed (\textit{e.g.} 32 for the file size below 10KB and 64 between 10KB and 30KB).
 Once the width of the image is fixed, we change its corresponding height depending on the file size.
 Figure~\ref{fig:fig1} illustrates the flowchart of malware transformation process with a binary file from the MS BIG database \cite{wang2015microsoft}, where a common Trojan horse download program (Dontovo A) is converted to an 8-bit grayscale image with the width of 64.
 Code segments of .text, .rdata, .data and .rsrc corresponds to 8-bit grayscale images are shown in Figure~\ref{fig:fig1}.

 \begin{figure*}[!htbp] 
    \centering
    \subfloat[Architecture of the proposed ATMPA Method]{\includegraphics[width=1.0\textwidth]{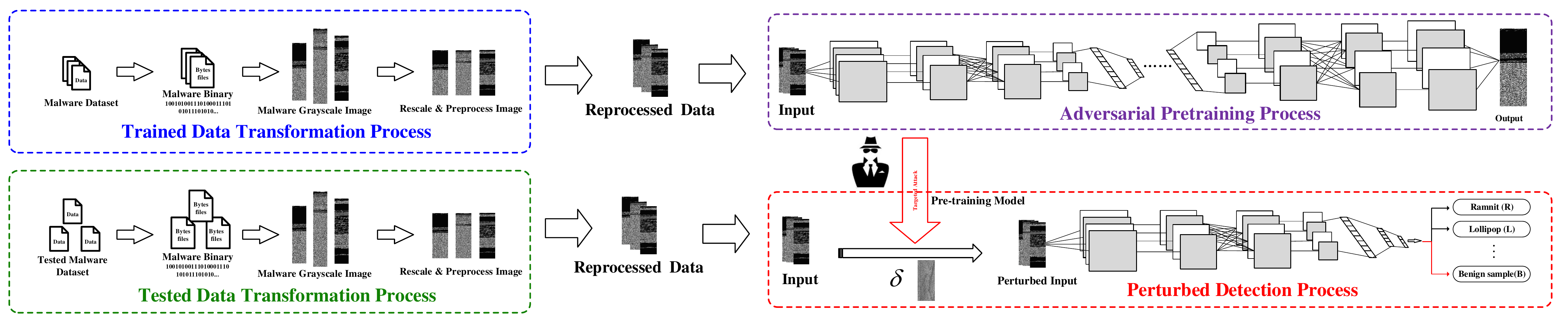}
    \label{fig:fig2a}}
    \hfil
    \subfloat[Architecture of the original Malware Detection Process]{\includegraphics[width=1.0\textwidth]{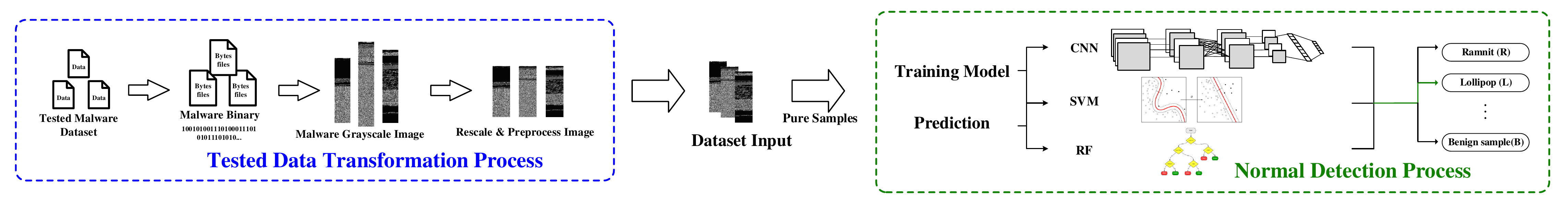}
    \label{fig:fig2b}}
    \hfil
    \caption{The Architecture of the ML-based Malware visualization Detection Process}
    \label{fig:fig2}
    \vspace{0.1in}
 \end{figure*}

 \subsection{ML-based Detection Methods}    
 \label{Pre MLDM}
 In a transformed malware image, the intensity, variance, skewness, kurtosis, and the average intensity value of each pixel are between 0 and 255.
 It is straightforward to extract feature information from the byteplot using Wavelet and Gabor-based methods, for example.
 After the feature extraction, the data set will be used to construct the discriminative model.
 In the following, three popular ML-based algorithms are discussed.

  \subsubsection{Random Forest (RF)}    
  \label{Pre MLDM RF}
  RF algorithm is an ensemble learning method for classification, regression and detection.
  RF constructs a multitude of decision trees at training time and produces the corresponding results according to the model classification and mean prediction of the individual trees \cite{ho1998random}.
  RF training algorithm uses bootstrap aggregating and bagging for tree learners.
  Predictions for unknown samples $x'$ are shown below by averaging the predictions from all the individual regression trees on $x'$.
   \begin{equation*}
    \begin{array}{l}
      $${\displaystyle {\hat {f}}={\frac {1}{B}}\sum _{b=1}^{B}f_{b}(x')}$$
    \end{array}
    \label{RFeq}
   \end{equation*}
  Alternatively, we can use the standard deviation of predictions to estimate the prediction uncertainty from all the individual regression trees on $x'$ by taking the majority vote in the case of classification trees, as
  \begin{equation*}
   \begin{array}{l}
      $${\displaystyle \sigma ={\sqrt {\frac {\sum _{b=1}^{B}(f_{b}(x')-{\hat {f}})^{2}}{B-1}}}}.$$
   \end{array}
  \end{equation*}

  \subsubsection{Support Vector Machine (SVM)}    
  \label{Pre MLDM SVM}
  SVM is a widely used machine learning algorithm \cite{cortes1995support} to fit a hyperplane to separate classes with the largest possible margin $M$.  
  In the SVM-based linear classification method, the unknown sample ($x$) can obtain a better classification effect in the generated hyperplane by optimizing the weight vector ($w$) and threshold ($\rho$) in decision function. The expression of assigned labels ($y$) is shown as follows,
  \begin{equation*}
   \begin{array}{l}
      $${y(x)=w^T x-\rho}$$
   \end{array}
  \end{equation*}

  Similarly, if the difference between malware images is too large that the dataset cannot fit with a linear model, we can simulate the dataset as a non-linear data distribution.
  Nonlinear decision functions are able to map samples into a feature space with special properties by applying nonlinear transformation.
  For the specificity of decision function, we use the $K(x, x_i)$ to represent the imaged sample, where $x$ denotes an unknown sample that needs to be compared and $x_i$ denotes the selected test sample in training dataset.
  Constructing a data pair for a parallel dataset is commonly referred to as a support vector and it can be expressed as:
  \begin{equation*}
   \begin{array}{l}
     $${\displaystyle {y(x)}=\sum _{\varepsilon \in SV}^{}\alpha_{i}y_{i}k(x,x_i)- \rho}$$.
   \end{array}
  \end{equation*}

  \subsubsection{Convolution al Neural Networks (CNNs)}    
  \label{Pre MLDM NN}
  An artificial neural network (ANN) is a computing system inspired by the biological neural networks \cite{mcculloch1943logical}. 
  ANNs with a single intermediate layer (\textit{i.e.} hidden layer) are qualified as shallow neural networks and some with multiple hidden layers are called as DNN, \textit{e.g.} CNN \cite{krizhevsky2012imagenet}.
  The basic structure of CNN consists of convolutional layers, pooling layers, fully connected layers and normalization layers.
  The feature information of malware is gradually enlarged by the extraction of multiple layers.
  CNN can be viewed as a mathematical model ${\displaystyle \textstyle F:x\rightarrow y}$, in which $F$ can be formalized as the composition of multidimensional and parameterized functions $f_i()$ corresponding to the layer of network architecture.
  Therefore, the formula $F$ can be expressed as:
  \begin{equation*}
   \begin{array}{l}
   $${\displaystyle \textstyle F:x\rightarrow f_n(...f_2(f_1(x,\theta_1),\theta_2)...,\theta_n)}$$,
   \end{array}
  \end{equation*}
  where a set of model parameters $\theta_i$ is learned during training.
  Each vector $\theta_i$ includes the weights for links to connect layer $i$ to layer $i-1$.
  Taking supervised learning as an example, parameter values can be estimated based on the prediction errors $f(x)-y$ through an input-output pair $(x,y)$.


\section{Methodology}
\label{Methodology}
Feature extraction and data reduction are commonly used in ML models, therefore malware may exist within the real world data samples due to the specificity and camouflage.
In particular, the introduction of AEs is of high possibility to produce the opposite result against the original results, which leads to severe security threat \cite{aeZhang2019}.
To demonstrate such vulnerability, we propose the ATMPA method to interfere with several ML-based visualization detectors.

 \subsection{Framework}
 \label{methodology framework}
 The attack model involved in this paper is that attackers can interfere with malware images during the visualization transformation without being identified by the original detectors, as shown in Figure~\ref{fig:fig2a}.
 As there exist some corresponding mechanisms \cite{lee2011study,nataraj2011malware,maiorca2017detection} between binary code and visualized images, the malware images can be converted back to the corresponding form as the binary code \cite{lee2011study, kim2018webmon}.
 AEs can also be converted back to the corresponding binary code through the reverse transformation.
 The framework of ATMPA consists of three functional modules: data transformation, adversarial pre-training and malware detection.
 Malware detection functions include AE generation and perturbed detection.
 The overall ATMPA framework is shown in Figure~\ref{fig:fig2a}.
 If there is no attack in the process (\textit{i.e.} no AE involved), the second module, adversarial pre-training, is not needed as shown in Figure~\ref{fig:fig2b}.
 Otherwise, the code segments of malware will be transformed into grayscale images in the data transformation module.
 We send malware images directly to ML-based malware detectors, \textit{i.e.} CNN, SVM and RF in this paper.
 The green detection arrow in Figure~\ref{fig:fig2b} indicates the normal detection process.

 When AE attack is in concern, we introduced a new module, adversarial pre-training.
 After a series of preprocessing including data transformation and normalization operations, the malware dataset required for AE generation is processed in the adversarial pre-training module and the dataset used for test will be propagated to malware detection module, as shown in Figure~\ref{fig:fig2a}.
 In the adversarial pre-training module, attackers can use machine learning methods to train an AE generation model with a known malware dataset, as shown Figure~\ref{fig:fig2a} (purpose block).
 Taking advantages of the AE generation model, attackers can produce a noise signal $\delta$ into the targeted samples.
 Attackers can apply small but intentionally worst-case perturbations on the transformed malware images to escape the observation from other users.

 When the original information has been obstructed, the targeted (or un-targeted) AEs coupled with the original dataset are propagated to the malware detector.
 As the original detectors are usually not able to resist AEs, the ML-based malware detectors will be induced to produce an incorrect result, as attackers desired.
 This is illustrated in the red arrow in Figure~\ref{fig:fig2a}.
 To evaluate the universal applicability of the ATMPA method, we choose the commonly used CNN-, SVM- and RF-based ML detectors.

 \subsection{Training process}
 \label{methodology TV}
 Since our proposed ATMPA method is not designed for a single detector, different types of detectors based on CNN, SVM, and RF are designed in the malware detection module.
 There are two training processes in this module: one is to achieve the function of normal malware detectors and the other is to obtain the AEs.
  \subsubsection{Pre-training process for malware detector}
  To obtain more accurate ML-based detectors, the pre-training process is usually performed with training samples to build a discriminant model.
  According to the discriminant model, the analyst can make a detection with the visualised malware images.
  As is shown in Figure~\ref{fig:fig2b}, different detectors in normal detection process have been pre-trained with the transformed malware dataset.
  This paper uses CNN, SVM and RF algorithms to build the corresponding malware detectors.

   \begin{figure}[b]
     \centering
     \includegraphics[scale=0.43]{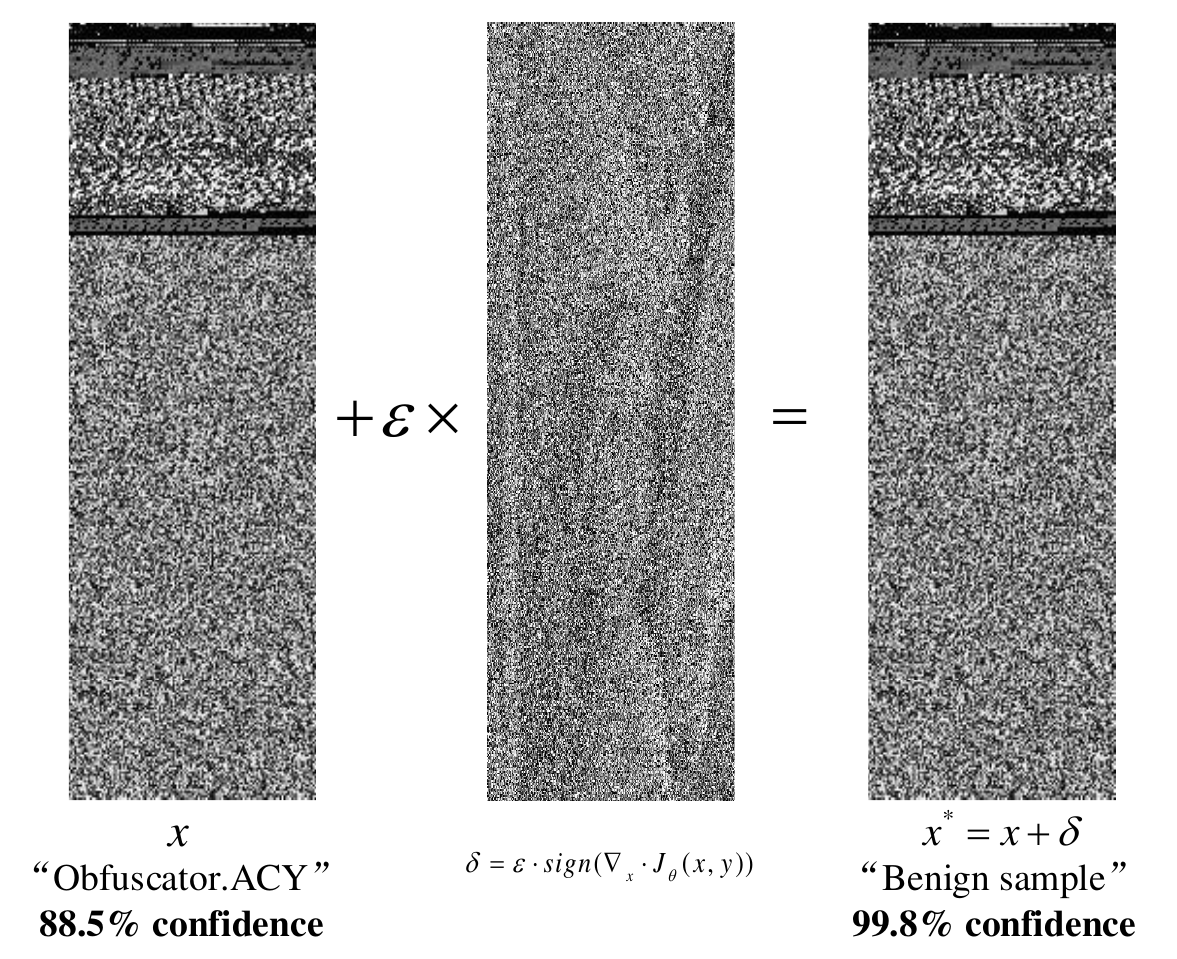}
     \caption{Process of Perturbed Visualized Malware}
     \label{fig:fig3}
   \end{figure}

  \subsubsection{Training process for adversarial pre-training}
  The proposed ATMPA uses GoogLeNet to generate malware AEs in the pre-training process.
  We use the binary indicator vector $X$ to represent malware samples without any particular structural properties or interdependencies.
  The feed-forward neural network is used in AE-training network. The rectifier is used as the activation function for each hidden neuron.
  Stochastic gradient descent and dropout are used for training.
  For normalization of the output probabilities, a softmax layer is employed as,
  \begin{equation*}
    \begin{array}{l}
      $$F_i(X)=\frac{e^{x_i}}{e^{x_0}+e^{x_1}}, x_i= \sum _{j{\mathop {=}}1}^{m_n}w_{j,i}\cdot x_j+b_{j,i}$$,
    \end{array}
  \end{equation*}
  where $F(\cdot)$ is an activation function, $x_i$ denotes to the $i_{th}$ input to a neuron and $w$ is the weight vector in gradient descent.

 There is no need for attackers to construct a new AE-training model in our proposed ATMPA method.
 AEs can be generated by using a similar CNN structure for malware detector training, therefore the generated AEs are difficult to detect and can be used to attack ML-based malware detectors.

 \subsection{Generation of Malware AEs}
 \label{methodology CAE}
  We define AEs as $x^*$ which are generated by introducing a subtle change to an original sample $x$ with $x^*=x+\delta$.
 In order to deceive the malware detector, two popular AE generation methods --- FGSM and C\&W attack --- are optimized and used in this paper.

  \subsubsection{FGSM-based method}
  \label{methodology CAE FGSM}
  FGSM \cite{goodfellow2014explaining} is a fast method to generate AEs that selects a perturbation parameter by differentiating this cost function with respect to the input itself.
  Along the direction of the gradient's sign at each pixel point, FGSM only performs one step gradient update.
  The perturbation is expressed as:
  \begin{equation}
  \begin{array}{l}
   $$\delta=\varepsilon\cdot sign(\nabla_{x}J_{\theta}(x,l))$$,
  \end{array}
  \label{Eq:eq1}
  \end{equation}
  where $\varepsilon$ is a small scalar value that restricts the magnitude of the perturbation, $sign(\cdot)$ denotes the sign function,$\nabla_{x} J_\theta(\cdot,\cdot)$ computes the gradient of the cost function $J$ around the current value $x$ with label $l$ of the model parameters $\theta$.


  In order to fool the classification better, an index $i$ with FGSM is exploited by finding the maximum gradient to replace the target class.
  \begin{equation*}
    \begin{array}{l}
      $$i=\underset{j}{\arg\max}(\nabla_{x_j}J_{\theta}(x_j,y^*))$$.
    \end{array}
  \end{equation*}

  We repeat this process with index($i$) until either the pre-judgment result $f(\cdot)$ producing the misclassification, \textit{i.e.,} $\underset{i}{\arg\max}(f(x^*)) = y^*$ or the index reaching the threshold of $i_{max}$.

  Taking the FGSM as an example, as shown in Figure~\ref{fig:fig3}, we illustrate the process of generating malware image (AEs) with subtle perturbation in details.
  The original malware sample can be recognized by the detector with 88.5\% confidence.
  After the perturbation with the distortion $\varepsilon=0.35$ through FGSM method, the perturbed malware samples induce the detector to output the benign sample with 99.8\% confidence.
  However, this subtle perturbation is difficult to observe by other users.


  \subsubsection{C\&W attack-based method}
  \label{methodology CAE CW}
  Carlini and Wagner \cite{carlini2017towards} summarized the distance metrics and proposed a new attack strategy with $L_0, L_2, L_\infty$ norm, named C\&W Attack.
  By optimizing the penalty term and distance metrics, attacking strategies driven by different $L_{p}$ norms have been expressed as follows:
  \begin{equation}
    \begin{array}{l}
     $$ \min\|\delta\|_{p}^2 = \min\|x^*-x\|_{p}^2 $$ \\
      \ \ \  $$~~s.t.~g(x+\delta)=y^* ~\&~x+\delta \in X $$ ,
    \end{array}
  \label{Eq:eq2}
  \end{equation}
  where $L_p$ norm penalty ensures the small quantity of added perturbation $\varepsilon$.
  The same optimization procedure with different \textit{L}-norm can conduct similar attacks with a modified constraint $g(x+\delta)\neq y$.

  Taking $l_2$ attack as an example, the perturbation $\delta$ is defined in terms of an auxiliary variable $w$
  \begin{equation*}
   \begin{array}{l}
     $$\delta=\frac{1}{2}(\tanh(w)+1)-x$$.
   \end{array}
  \end{equation*}
  Then, we optimize $w$ to find perturbation factor $\delta$ by
  \begin{equation}
    \begin{array}{l}
      $$\underset{w}{\min}\|\frac{1}{2}(\tanh(w)+1)\|_2+c\cdot g(\frac{1}{2}(\tanh(w)+1))$$,
    \end{array}
    \label{Eq:eq3}
  \end{equation}
  where $g(\cdot)$ is an objective function based on the hinge loss,
  \begin{equation*}
    \begin{array}{l}
     $$ g(x^*)=\max(\underset{\delta}{\max}(Z(x^*)_i)-Z(x^*)_t,-\kappa)$$.
    \end{array}
  \end{equation*}
  $Z(\cdot)_i$ represents the softmax function for class $i$, $t$ is the target class and $\kappa$ denotes a constant to control the confidence with which the misclassification occurs.


  In the proposed ATMPA method, $L_p$-based C\&W attack method is also introduced to generate malware AEs, including the $l_0,\ l_2\ \&\ l_\infty$ attack.
  Considering the detector's hierarchical structure in malware images, the optimization formulation and optimized AE generating process are modified in the ATMPA method according to \cite{carlini2017magnet}.
  Take $l_2$ attack for instance, we optimise the formula in Eq.~\ref{Eq:eq2} into
  \begin{equation}
    \begin{array}{l}
     $$\min\|x^{*}-x\|_{2}^{2},\  \\
        \ \ s.t.\ f(x^*)\neq f(x)$$.
    \end{array}
  \label{Eq:eq4}
  \end{equation}
  where the original single optimization item $c\cdot g(x^*)$ from the formula $\min\|x^{*}-x\|_{2}^{2}+c\cdot g(x^*)$ in Eq.~\ref{Eq:eq3} are divided into two parts as $R_i$ \& $D_i$ in the Eq.~\ref{Eq:eq4}.
  $R_i=r\cdot g_r(x^*)$ and $D_i=d\cdot g_d(x^*)$ are expressed as $l_r(x^*)$ and $l_d(x^*)$, denoting the corresponding loss function in the classifier and detector.
  The $r$ and $d$ are chosen via binary search.


\section{Experiments}
\label{Experiment}

 \subsection{Dataset}
 \label{Experiment Data}
 To evaluate the performance of the proposed method, we use an open source malware dataset in Kaggle Microsoft Malware Classification Challenge (BIG 2015)~\cite{wang2015microsoft}.
 It consists of 10,867 labeled malware samples with nine classes.
 To collect benign examples, we scrap all the valid executables on Windows 7 and XP on the virtual machine.  
 We select 1000 benign samples by using anti-virus vendors in VirusTotal intelligence search\footnote{https://www.virustotal.com/en/}.
 The distribution of these samples is illustrated in Table~\ref{table:tab1}.
 Considering the robustness and scalability of the structure in neural networks, we adopt the state-of-the-art GoogleNet Inception V3 architecture for AE generation.
 Finally, the FGSM and C\&W attack methods are used to generate AEs ($x^*$) with the perturbation ($\delta$).
 In this paper, the generated AEs belongs to targeted AEs, called as pseudo-benign samples, able to disguise malware to deceive the detector.

 \begin{table}[t]
     \centering
     \caption{The MS BIG dataset with malware class types and benign samples}
     \scalebox{0.8}{
     \begin{tabular}{lc}
      \toprule
      Malware types  & Number of Samples \\
      \midrule
      Ramnit (R)         & 1534  \\
      Lollipop (L)       & 2470  \\
      Kelihos ver3 (K3)  & 2942  \\
      Vundo (V)          & 451   \\
      Simda (S)          & 41    \\
      Tracur (T)         & 685   \\
      Kelihos ver1 (K1)  & 386   \\
      Obfuscator.ACY(O)  & 1158  \\
      Gatak (G)          & 1011  \\
      Benign sample (B)   & 1000  \\
     \bottomrule
     \end{tabular}}
     \label{table:tab1}
 \end{table}

  \subsection{Attack Effectiveness}
 \label{Experiment AER}
 The effectiveness of proposed ATMPA is evaluated by attacking three commonly used malware detectors based on CNN, SVM and RF algorithms.
 Through the method of 10-fold cross-validation, the original malware sample set is randomly partitioned into ten equal sized subsample sets, with one set of AE subsamples is replaced.
 Different AEs can be generated by using FGSM and C\&W attack method with different distortion parameter $\varepsilon$.
 The experimental results demonstrate that all three ML-based detectors are easily misled.

 \begin{table}[b]      
    \centering
    \caption{Attack results with FGSM method}
    \scalebox{1.0}{
    \begin{tabular}{cccc}
      \toprule
      \multicolumn{2}{c}{\multirow{1}{*}{Detection Methods}}
      &\multicolumn{1}{c}{\multirow{1}{*}{Distortion ($\varepsilon$)}}
      &\multicolumn{1}{c}{\multirow{1}{*}{Success rate (\%)}}
      \\
      \midrule
      \multicolumn{1}{c}{\multirow{2}{*}{CNN}}& Basic CNN & 0.4 & 100\\
      \multicolumn{1}{c}{}& Wavelet+CNN                   & 0.5 & 100\\
      \cline{2-4}
      \multicolumn{1}{c}{\multirow{2}{*}{SVM}}& Basic SVM & 0.5 & 100\\
      \multicolumn{1}{c}{}& Wavelet+SVM                   & 0.6 & 99.8\\
      \cline{2-4}
      \multicolumn{1}{c}{\multirow{2}{*}{RF}} & Basic RF  & 0.4 & 100\\
      \multicolumn{1}{c}{}& Wavelet+RF                    & 0.6 & 99.7\\
      \bottomrule
    \end{tabular}}
    \label{table:tab2}
 \end{table}

 \begin{figure*}[!htbp] 
    \centering
    \subfloat[Three basic learning methods]{\includegraphics[width=0.32\textwidth]{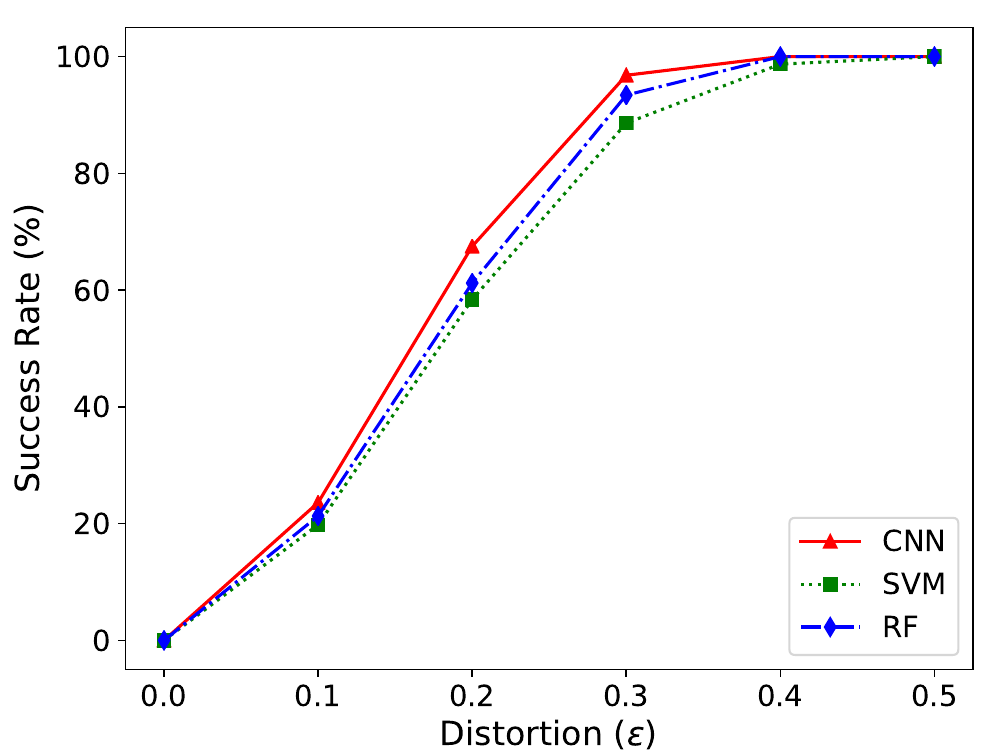}
    \label{fig:fig4a}}
    \hfil
    \subfloat[Three wavelet combined methods]{\includegraphics[width=0.32\textwidth]{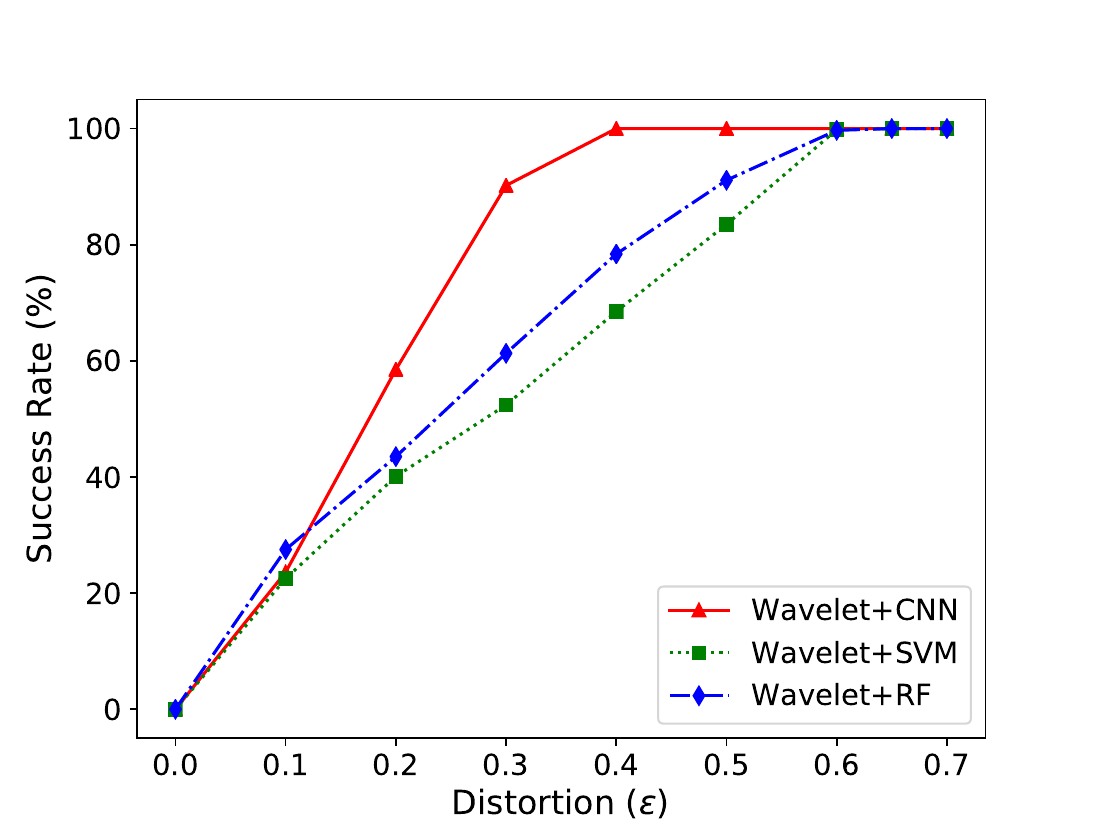}
    \label{fig:fig4b}}
    \hfil
    \subfloat[All tested methods]{\includegraphics[width=0.32\textwidth]{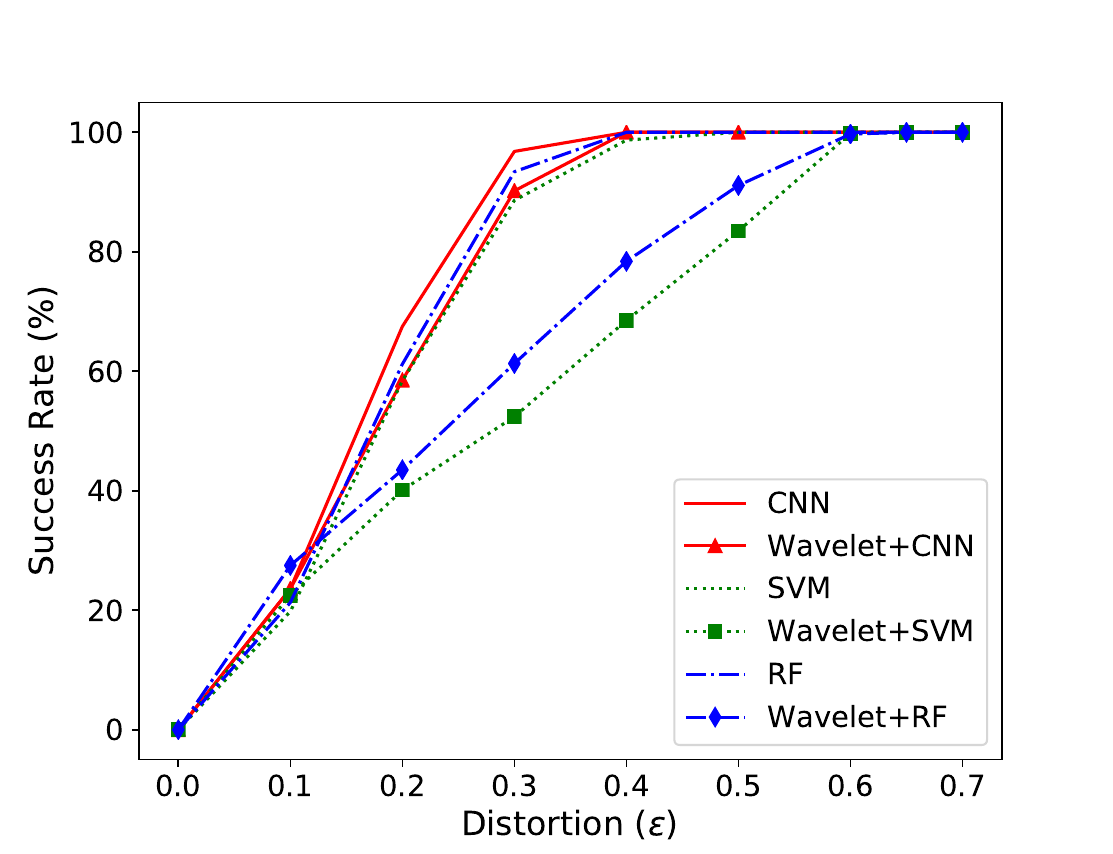}
    \label{fig:fig4c}}
    \hfil
    \caption{Comparison analysis for different malware detectors with FGSM attack}
    \label{fig:fig4}
    \vspace{0.1in}
 \end{figure*}

 \subsubsection{CNN}
  \label{Experiment AER CNN}
  To begin with, we use the pseudo-benign AEs to attack a CNN visualization detector built on GoogleNet Inception V3.
  We adjust the default distortion parameter $\varepsilon$ around a default value for 10 iterations of the gradient descent direction.
  A CNN-based detector is easily induced and produces wrong results.
  Through the analysis of cross-validation, we find that if the value of the distortion parameter $\varepsilon$ is adjusted to $\varepsilon=0.4$, the success rate of attacks is 100\%, as shown in Table~\ref{table:tab2}.
  The success rate grows as the corresponding increase of distortion value ($\varepsilon$) which is depicted in Figure~\ref{fig:fig4}.

   If a small learning rate, such as 0.05, is assigned to prevent the feature information of original sample from fast extraction, AEs generated from C\&W attack easily fool the CNN-based detectors after 100 iterations of gradient descent direction.
  In Table~\ref{table:tab3}, we list the success rate with different distortion values in different \textit{L}-norms distance to express the high success rate in C\&W attack.
  Furthermore, cases with interference parameters as 0.2 to 0.4 achieve a success rate with 100\% for all the attacks on the CNN-based detector.

  \subsubsection{SVM}
  \label{Experiment AER SVM}
  Because of the different mechanisms between SVM and CNN, SVM-based malware detector is attacked to verify the scalability of the proposed ATMPA.
  An SVM-based malware detector with the default parameters ($\gamma=0.1$, $C=10^2$ and $k=10$) is used in this experiment. 
  As it shows in Table~\ref{table:tab2}, when the value of distortion $\varepsilon$ has been set to 0.5, the attackers can obtain a success rate of 100\%.
  Comparing to other baseline detection methods, the SVM-based detector is difficult to resist the attack from ATMPA, though its distortion is the biggest.
  Meanwhile, as shown in Figure~\ref{fig:fig4a} with the blue dashed line, FGSM-based attack can achieve up to 100\% success rate with the distortion increasing.
  In particular, when the value of distortion ($\varepsilon$) is over 0.4, the corresponding SVM-based detector will become invalid.

  \subsubsection{RF}
  \label{Experiment AER RF}
  Apart from CNN and SVM-based detectors, we also use the pseudo-benign AEs to attack the RF-based detector with the default parameter settings ($\textnormal{mtree}=1000, \textnormal{ntree}=100$). 
  By using statistical evaluation methods in 10-fold cross-validation, the mean success rate could be easy to reach 99.7\% if the distortion value set to 0.6 in FGSM method.
  In addition, the similar success rate is reached when C\&W attack-based method is used, as it listed in Table~\ref{table:tab3}.
  If attackers set the distortion value over 0.36, all of the \textit{L}-norm attacks (including $l_0,l_2~\&~l_\infty$) can successfully induce the RF-based detector.
  Therefore, ATMPA can fool RF-based detectors effectively.

 \begin{table}[b]
    \centering
    \caption{Attacking results with C\&W attack method}
    \captionsetup{font={small}}
    \scalebox{0.9}{
    \begin{tabular}{cccccc}
      \toprule
      \multicolumn{2}{c}{\multirow{2}{*}{Detection Methods}}
      &\multicolumn{3}{c}{\multirow{1}{*}{Distortion}}
      &\multicolumn{1}{c}{\multirow{2}{*}{Success rate (\%)}}
      \\
      \cline{3-5}
      \multicolumn{2}{c}{}
      &\multicolumn{1}{c}{$L_0$} & {$L_2$} & {$L_\infty$}
      &\multicolumn{1}{c}{}
      \\
      \midrule
      \multicolumn{1}{c}{\multirow{2}{*}{CNN}}& Basic CNN & 0.23 & 0.30 & 0.36 & 100\\
      \multicolumn{1}{c}{}& Wavelet+CNN                   & 0.34 & 0.41 & 0.40 & 100\\
      \cline{2-6}
      \multicolumn{1}{c}{\multirow{2}{*}{SVM}}& Basic SVM & 0.51 & 0.43 & 0.40 & 100\\
      \multicolumn{1}{c}{}& Wavelet+SVM                   & 0.69 & 0.65 & 0.32 & 99.7\\
      \cline{2-6}
      \multicolumn{1}{c}{\multirow{2}{*}{RF}} & Basic RF  & 0.49 & 0.43 & 0.39 & 100\\
      \multicolumn{1}{c}{}& Wavelet+RF                    & 0.56 & 0.52 & 0.49 & 100\\
      \bottomrule
    \end{tabular}}
    \label{table:tab3}
 \end{table}

  \subsubsection{Comparisons}
  Experimental results show that the robustness of anti-AE ability in CNN-based detectors is the most vulnerable to ATMPA.
  SVM-based detector performs better in resisting adversarial attacks than CNN- and RF-based detectors, as demonstrated with higher distortion in Table~\ref{table:tab2}.
  Analysis is made in terms of the curvature and gradient of these three curves in Figure~\ref{fig:fig4a}, where the curve of SVM-based detector illustrates the best robustness, even though its ability of anti-AE attack is limited.
  With a slight adjustment of distortion scale in the FGSM and C\&W attack method, ATMPA can easily break through the defence of the SVM-based detector.
  The mean value of distortion from cross-validation analysis in SVM-based detectors is larger than others indicating a good model transferability of ATMPA.
  Our proposed attack method can hence be extended to liner-based surprised learning methods.	
  With respect to the RF-based detectors, ATMPA shows similar features for RF- and CNN-based detectors, as illustrated in Table~\ref{table:tab2} and Figure~\ref{fig:fig4a}.
  A similar increasing trend is observed for the curvature of the whole increased curve.
  When the attacking efficiency reaches the full attack (\textit{i.e.,} success rate is 100\%), the distortion values $\varepsilon$ for AE-generation are close to the factor of 0.5.
  Therefore, the ability to defend ATMPA in RF-based malware detectors is very limited compared to the SVM counterpart.

 \subsection{Attack Transferability }
 \label{Experiment ATR}
 Transferability comparison is made from two aspects.
 Firstly, attacks using similar detection methods defending we attack the similar detectors such as CNN and wavelet-combined CNN detectors using the same AE set from the ATMPA,.
 From the perspective of different types of detection methods, we also use the same set of AEs from the ATMPA, but attack malware detectors using CNN, SVM and wavelet-combined SVM \textit{etc.}
 In this section, the wavelet is used to strengthen the robustness on ML-based detection methods.
 Wavelet transformation, a classical technique of feature extraction and spatial compression \cite{lin2000feature}, has been applied in many research fields and wavelet-combined ML methods have been widely used in the field of image classification.
 If the malware image is processed through the wavelet transformation, the sparse space of its own image will be compressed.
 We use the same group of AEs to attack CNN-, SVM- and RF-based malware detectors optimized by the wavelet technique. 
 As expected, the generated AEs can also induce the optimized detector.
 The comparison results in terms of transferability are shown in Table~\ref{table:tab4}.
 \begin{table}[b]   
     \centering
     \captionsetup{font={small}}       
     \caption{Transferability comparison in different detectors}
     \scalebox{0.76}{
     \begin{tabular}{ccccccc}
      \toprule
          & CNN  & SVM & RF & Wavelet-CNN  & Wavelet-SVM & Wavelet-RF \\
      \midrule
      CNN & 100\%  & 57.8\% & 86.3\% & 88.5\% & 54.2\% & 75.6\% \\  
      SVM & 64.5\% & 100\%  & 75.5\% & 54.2\% & 85.7\% & 64.5\% \\  
      RF  & 81.5\% & 45.6\% & 100\%  & 67.9\% & 48.2\% & 88.7\% \\
     \bottomrule
     \end{tabular}}
     \label{table:tab4}
 \end{table}

  \subsubsection{Wavelet-CNN}
  \label{Experiment ATR WCNN}
  To attack the wavelet-combined CNN detector (Wavelet-CNN), ATMPA can defeat the detector with a strong transferability.
  By increasing the range of the distortion value $\varepsilon$, Wavelet-CNN detector can be induced to produce the malware as a benign label.
  The error rate of the detector will rise from 85\% to 100\% through increasing the intensity of the distortion degree in FGSM method.
  Similarly, the Wavelet-CNN detector will also produce an error rate of 100\% when using C\&W attack.
  Attackers only need to increase the distortion range from 0.15 to 0.18 on different \textit{L}-norm distances.
  Generally, AEs can successfully fool the CNN-based detectors in accordance with the attacker's pre-set label $y^*$, when the distortion value is more than 0.5.
  Whether FGSM or C\&W attack is used, the generated AEs ($S_{AE}$) can achieve a 100\% success rate in terms of attacking the CNN-based and wavelet-combined detectors.
  $S_{AE}$ is used to test the transferability by attacking different detectors.
  Experiment results show that the maximum transferability achieved is up to 88.5\%.

  \subsubsection{Wavelet-SVM}
  \label{Experiment ATR WSVM}
  Wavelet-combined SVM (Wavelet-SVM) detection is used to measure the transferability of the ATMPA method.
  Wavelet-SVM malware detector is more resistant than Wavelet-CNN methods, but it is difficult to resist the attack from ATMPA.
  With a fixed distortion parameter as $\varepsilon =0.4$ and $\varepsilon=0.5$, Wavelet-SVM shows the lowest success rate.
  The ability to resist AEs has been greatly improved in the SVM-based detectors.
  As for the comparison between Wavelet-SVM and SVM-based detectors, Wavelet-SVM can resist $20\%\thicksim30\%$ AE attacks, even though such AEs can fool the SVM-based detector completely.
  The transfer rate in different detectors is shown in Table~\ref{table:tab4} by using the same AEs.
  When the distortion ($\varepsilon$) is 0.3 in FGSM, or the \textit{L}-norm distance is 0.71 in C\&W attack, the AE resistant SVM detector will be attacked effectively.
  Therefore, even though the wavelet-SVM detector is more resistant to AEs, attackers can also fully induce these detectors by choosing appropriate perturbations.

  \subsubsection{Wavelet-RF}
  \label{Experiment ATR WRF}
  As shown in Table~\ref{table:tab4}, the maximum transfer rate for RF and wavelet-RF detectors is 88.7\%.
  Parameters are assigned consistently as mentioned above ($\textnormal{mtree}=1000$, $\textnormal{ntree}=100$).
  Wavelet-RF detection method can express a certain robustness when resisting the attack from AEs, but the defensive ability is extremely limited.
  Taking the FGSM method as an example, when the distortion$\varepsilon$ is 0.4, the success rate of attacking the RF detector is 100\%.
  If the distortion increases 15\%, the attack on wavelet-RF will achieve the success rate of 100\% as well.
  As for C\&W attack methods, it does not produce a very large fluctuation even when wavelet-combined technique is introduced.
  Through increasing the distance of $L_1, L_2, L_\infty$ to values of 0.45, 0.55, 0.65 respectively, the pseudo-benign AEs generated by the C\&W attack method can achieve 100\% success rate.
  Therefore, with a better transferability than FGSM, the C\&W attack-based ATMPA will be easier to attack RF-based detectors.

\section{Defensive strategy}
\label{Discussion}
By analyzing the differences between different AEs and original samples, the iterative process of generating AEs and the algorithm structure on different detectors, we summarise some qualitative defensive strategies to defend AE attack.

At first, we compare the mean distance $D_x^*$ and $D_x^t$ between the original inputs ($x$) and corresponding normal AE samples ($x^*$) and pseudo-benign AE samples $x^t$.
We find that both $D_x^*$ and $D_x^t$ have similar values, with the proportion of $1$:$1.08$.
The normal AEs ($x^*$) generated for obstructing malware detectors perform similarly to pseudo-benign AEs ($x^t$) and its corresponding mathematical distribution is close to the uniform distribution, which expresses the universality and expansibility in the process of generating AE to confuse the adversary.
Therefore, with the subtle perturbation, it is easy for AEs to fool the detectors.

\begin{figure}[b]   
  \centering
  \includegraphics[scale=0.48]{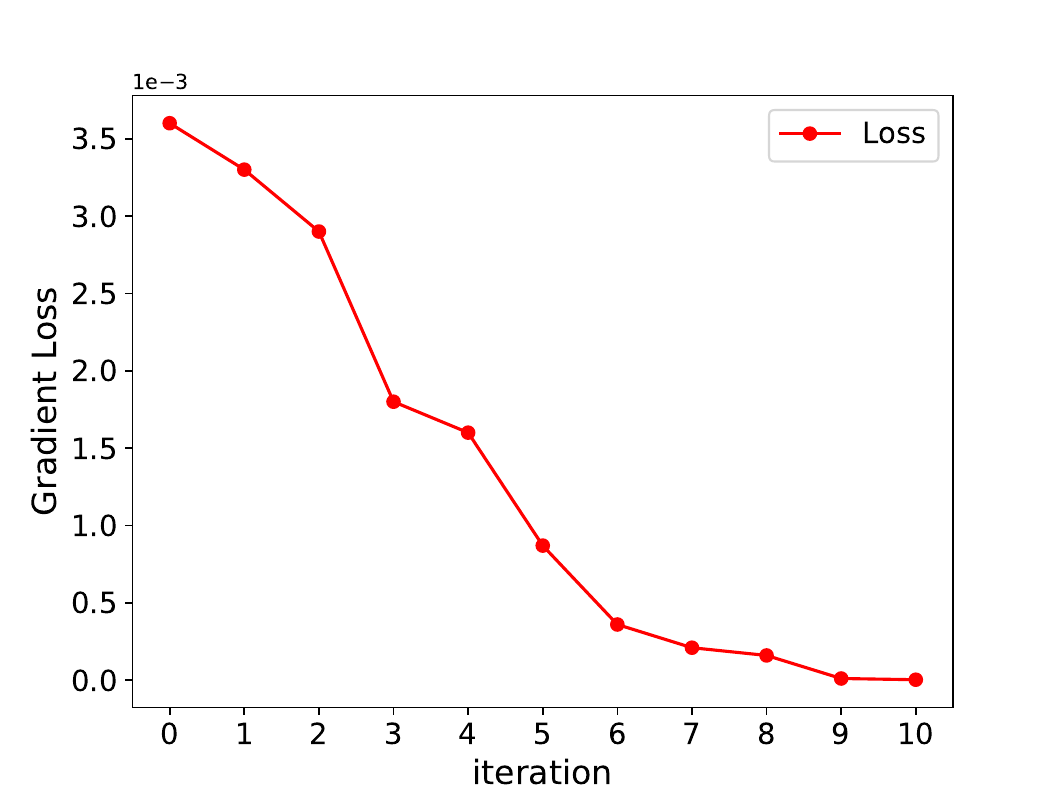}
  \caption{The Loss value with iteration}
  \label{fig:fig5}
\end{figure}

The process of AE generation fits along the optimal gradient of variants in the target category. 
Taking FGSM method as an example, Figure~\ref{fig:fig5} shows the variant trend of the absolute value ($|l|$) of gradient loss ($l$) in the objective function $f(\cdot)$ with the number of iterations ($t$) growing.
For the target sample ($x$) of the first iteration, the gradient loss ($l$) of the sample ($x$) has the largest value (as $l_0$=3.61$\times10^{-3}$).
With the subtle perturbations in each variables along the direction of gradient descent, the value of gradient loss ($l$) will rapidly reduce to zero, which shows a reduction in magnitude of two orders.
The entire process of AE generation can be considered as a learning stage for target sample ($x$).
Therefore, the malware visualization detection method based on level-by-level learning methods is difficult to resist the attack from AEs.

Since AEs used in this paper are based on CNN with a hierarchical structure, one can find that there exist some differences among other ML algorithms with linear structures such as SVM.
That is why the SVM and wavelet-SVM based malware detectors can resist this kind of AE attack to some extent.
In contrast, the RF and CNN-based detectors with similar hierarchical discriminant structure are vulnerable to AE attacks.
We can utilize this feature to resist the AE attack.
When we use the ATMPA method to attack the detectors with similar algorithmic structures, their transferability rates are usually higher than those with different algorithmic structures, as listed in Table~\ref{table:tab4}.
For example, the transfer rate from CNN to RF is 86.3\%, but from RF to SVM, the rate reduces to 45.6\%.
Therefore, the AEs generated by using NNs can not only make the conventional detectors fail but also attack other algorithms in similar structure with a high transferability.

Therefore, it is difficult to defend the AE attack for conventional NNs-based malware detectors.
However, some possible defensive strategies are proposed against the AE attack based on the above observations.
\begin{itemize}
\item According to the similarity among different AEs and distribution features among different perturbation ($\delta$), the accuracy can be improved if the dataset can be preprocessed to expand the difference between samples.
\item According to the generation process of AEs including the variation of gradient loss ($l$) and iteration number ($t$), the feature information of the normal samples could avoid to be learned by the attackers through the process of expanding the dataset or improving their differences.
\item According to the characteristic of different algorithmic structures in malware detectors and AE generation process, it would be useful that defenders can build a security module to resist the perturbation from AEs in ML-based malware visualization detection. For example, the linear optimization detection based on Linear Discriminant Analysis (LDA), Logistic Regression and Hyperplane Separating method can be used.
\end{itemize}

\section{Conclusion}
\label{Conclusion}
In this paper, we propose the first attack method, ATMPA, to fool visualized malware detectors.
The ATMPA method uses the gradient-based FGSM method and \textit{L}-norm based C\&W attack method to generate AEs on the converted image dataset.
The tested ML-based visualization detectors are all unable to detect the malware correctly.
Experimental results demonstrate that very weak adversarial noise can achieve 100\% successful attack rate for CNN, SVM and RF-based malware detectors.
In addition, the transfer rate when attacking different malware detectors is up to 88.7\%.

In future, a new malware defense method using AEs could be designed to simulate the variant of malware.
In combination with adversarial training or Generative Adversarial Networks (GAN), it will increase the ability of defense against zero-day attacks.
Moreover, AEs are trained per classification problem, meaning that they can operate on the set of labels that have been trained.
Switching to an alternative set of labels is likely to reduce their attacking effectiveness.
Finally, the ATMPA can be used for other modalities, such as sound/speech processing to address users with recognition impairments.




\begin{acks}
This work is supported by the National Natural Science Foundation of China (Grant No. 61874042, No. 61602107, No.61472125), the Hu-Xiang Youth Talent Program (Grant No. 2018RS3041), the Natural Science Foundation of Hunan Province, China (Grant No. 618JJ3072), and the Fundamental Research Funds for the Central Universities.
\end{acks}

\bibliography{ATMPA}
\bibliographystyle{ACM-Reference-Format}





\end{document}